\documentclass[twocolumn,showpacs,amsmath,amstex,amssymb,mathfonts,prl]{revtex4-1}
\usepackage{graphicx,bm,units}

\begin{document}

\newcommand{\vk}{{\bf k}}
\def\ns{^{\vphantom{*}}}
\def\ket#1{{|  #1 \rangle}}
\def\bra#1{{\langle #1|}}
\def\braket#1#2{{\langle #1  |   #2 \rangle}}
\def\expect#1#2#3{{\langle #1 |   #2  |  #3 \rangle}}
\def\cH{{\cal H}}
\def\half{\frac{1}{2}}
\def\sut{\textsf{SU}(2)}
\def\suto{\textsf{SU}(2)\ns_1}
\def\kF{\ket{\,{\rm F}\,}}

\title{Robust extended states in a topological bulk model with even spin-Chern invariant}

\author{Hadassah Shulman and Emil Prodan}
\address{Department of Physics, Yeshiva University, New York, NY 10016} 

\begin{abstract} This paper demonstrates the existence of topological models with gapped edge states but protected extended bulk states against disorder. Such systems will be labeled as trivial by the current classification of topological insulators. Our finding calls for a re-examination of the definition of a topological insulator. The analysis is supported by extensive numerical data for a model of non-interacting electrons in the presence of strong disorder. In the clean limit, the model displays a topological insulating phase with spin-Chern number $C_s$=2 and gapped edge states. In the presence of disorder, level statistics on energy spectrum reveals regions of extended states displaying levitation and pair annihilation. Therefore, the extended states carry a topological invariant robust against disorder. By driving the Fermi level over the mobility edges, it is shown that this invariant is precisely the spin-Chern number. The protection mechanism for the extended state is explained.
\end{abstract}


\date{\today}

\maketitle

According to a widely accepted definition \cite{Kane:2006xu}, a topological insulator is a material that does not conduct electricity in the bulk but displays dissipationless conducting channels at the edges. This definition has been recently debated based on new topological models \cite{Turner2010cu,Hughes2010gh} that display gapped edge states and yet they cannot be connected to a trivial insulator without closing the insulating gap or breaking the symmetry that defines them. These new models display anomalous responses and properties that are topologically protected, but they don't fit in the current universal classification scheme \cite{Schnyder:2008qy}. The problem of classifying the insulators has been reopened.  

The present paper gives hard evidence that indeed, the criterium based solely on the edge states is too restrictive, and that large classes of materials with interesting and potentially useful protected properties can be neglected because of that. We devised a 2-dimensional (2D) lattice model that has gapped edge states but protected extended bulk states against disorder. The model is based on the Kane-Mele Hamiltonian for graphene \cite{Kane:2005np,Kane:2005zw},
\begin{equation}\label{KaneMele}
\begin{array}{c}
H_0^{\mbox{\tiny{QSH}}}=\sum\limits_{\langle {\bm m \bm n} \rangle,\sigma} |{\bm m},\sigma\rangle \langle {\bm n},\sigma| \medskip \\
+\sum\limits_{\langle \langle {\bm m \bm n} \rangle \rangle,\sigma}  \alpha_{\bm n}(t/2+i\eta [\hat{ {\bm s}} \cdot \underline{{\bf d}_{\bm k \bm m}\times{\bf d}_{\bm n \bm k} }]_{\sigma,\sigma} )|{\bm  m},\sigma\rangle \langle {\bm  n},\sigma| \medskip \\
+i\lambda\sum\limits_{\langle {\bm m \bm n} \rangle,\sigma \sigma'}  [ {\bf e}_z\cdot (\hat{{\bf s}}\times \underline{{\bf d}_{\bm m \bm n}})]_{\sigma,\sigma'} |{\bm m},\sigma\rangle \langle {\bm n},\sigma'|,
\end{array}\nonumber
\end{equation}
but we use spin operators $\hat{\bm s}$ appropriate for spin $\frac{3}{2}$ particles (thus $\sigma=\pm\frac{1}{2},\pm\frac{3}{2}$). The notation in Eq.~\ref{KaneMele} is explained in Ref.~\cite{ProdanJPhysA2010xk}. The  resulting model is suited for the present analysis because a) we can compare with the original Kane-Mele model, b) its bands are highly entangled so we are not dealing with just two copies of the original Kane-Mele model, and c) the model remains time-reversal invariant. The original and the new model will be referenced as the spin $\frac{1}{2}$ and $\frac{3}{2}$ models, respectively.

\begin{figure}
  \includegraphics[width=8cm]{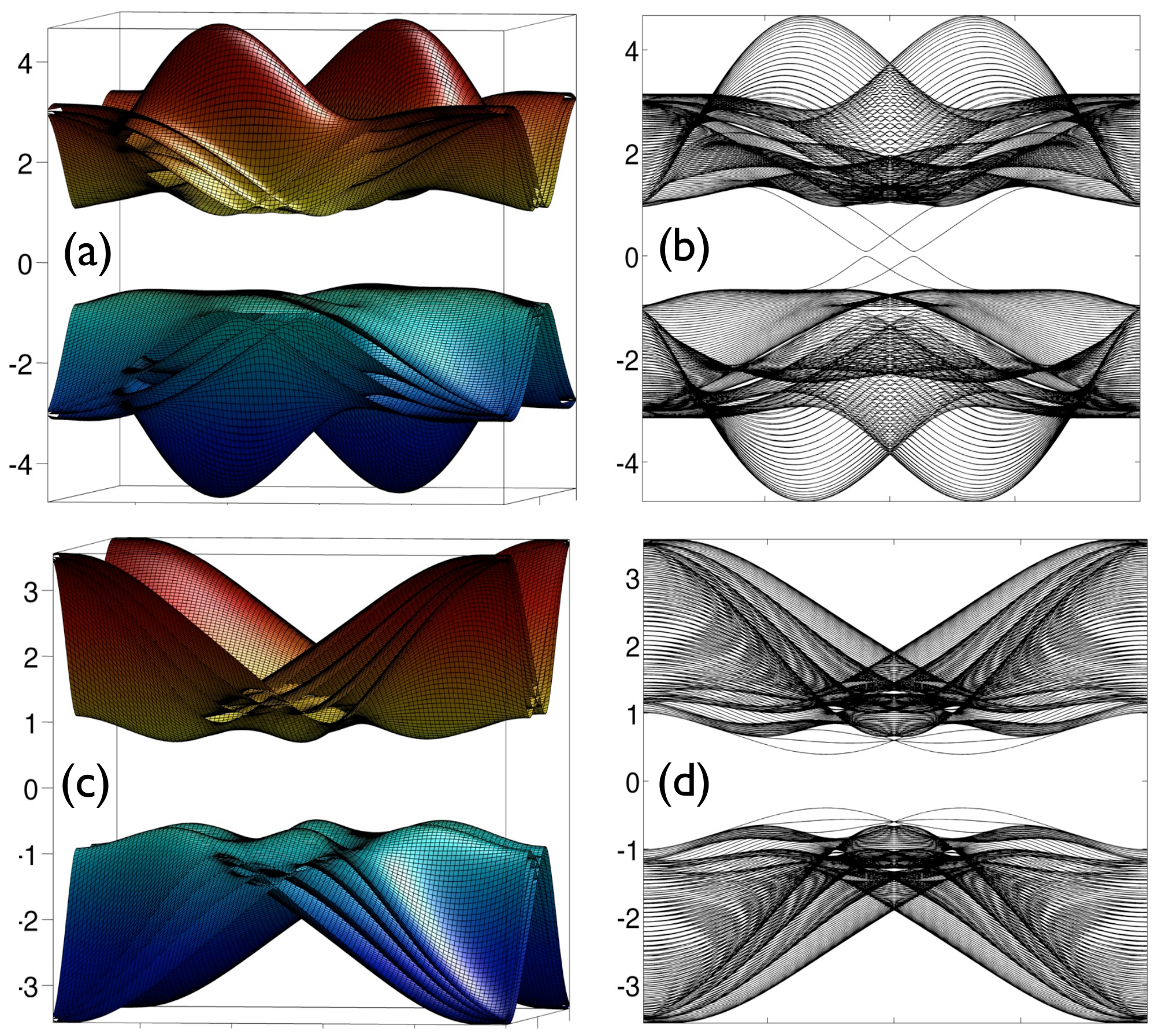}\\
  \caption{The bulk (left) and ribbon (right) bands of the spin $\frac{3}{2}$ model for the $C_s$=2 case (upper) with $t$=0, $\eta$=0.6 and $\lambda$=0.3 and for the $C_s$=0 case (lower) with $t$=0.6, $\eta$=0 and $\lambda$=0.3.}
 \label{CleanSystem}
\end{figure}

In the clean limit, the spin $\frac{3}{2}$ model displays eight bands separated by an insulating gap, which closes for exceptional values of the parameters $(t,\eta,\lambda)$ on a surface that separates the 3D parameter space into distinct regions, among which one with $C_s$=2 and one with $C_s$=0. For example, $C_s$=2 for $t$=0, $\eta$=0.6 and $\lambda$=0.3. A plot of the bulk bands for this case is shown in Fig.~\ref{CleanSystem}(a). The bands for a ribbon geometry with open boundary conditions, plotted against the conserved momentum, is shown in Fig.~\ref{CleanSystem}(b). There are four edge bands emerging from the bulk spectrum, which cross each other at $k$ points where the Kramer's degeneracy is not protected, so the bands hybridize at these crossings and become gapped. Note that the Rashba term responsible for this hybridization is large (the avoided crossings occurs far away from $k$=0) yet the gap is small. If we choose $t$=0.6, $\eta$=0 and $\lambda$=0.3, then $C_s$=0. A plot of the bulk and ribbon bands are shown in Figs~\ref{CleanSystem}(c-d). The ribbon bands display an almost clean bulk gap, except for accidental edge bands lying very close to the edges of the bulk spectrum.

\begin{figure}
  \includegraphics[width=7cm]{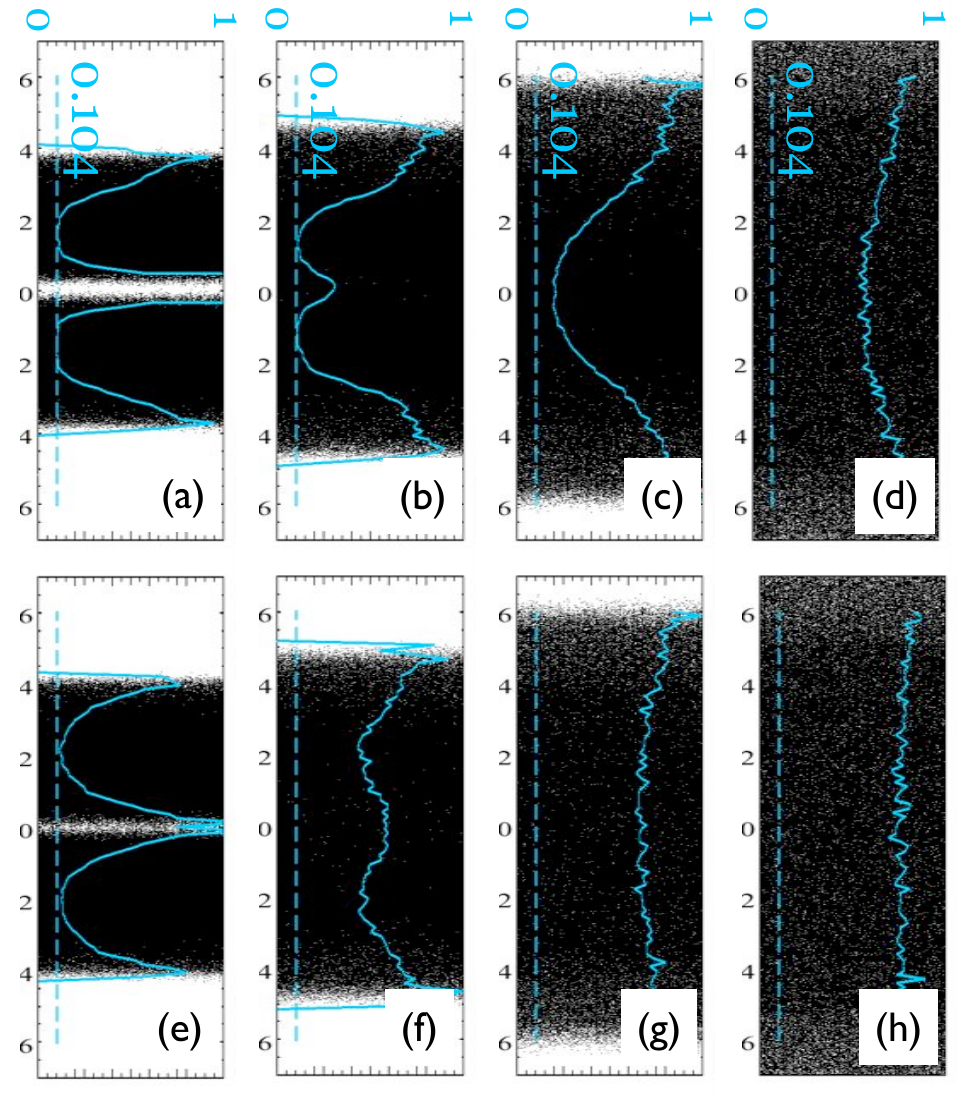}\\
  \caption{Energy spectrum and the variance of the level spacings for the spin $\frac{1}{2}$ model. The top corresponds to the topological case $t$=0, $\eta$=0.6 and $\lambda$=0.3 with $C_s$=1 and the bottom to the trivial case $t$=0.6, $\eta$=0 and $\lambda$=0.3 with $C_s$=0.}
 \label{SpinChern1LevelStatistics}
\end{figure}

We now add spin-indepedent disorder: 
\begin{equation}
H_0 \rightarrow H_\omega = H_0 + W\sum_{{\bm n},\sigma} \omega_{{\bm n}} |{\bm n},\sigma\rangle \langle {\bm n},\sigma|,
\end{equation}
where $\omega_{\bm n}$ are random entries uniformly distributed between $-\frac{1}{2}$ and $\frac{1}{2}$. The amplitudes $\omega_{{\bm n}}$ are the same for all the states $|{\bm n},\sigma\rangle$ in a unit cell. We use the level statistics analysis to probe the localized/delocalized character of the bulk quantum states, which was previously demonstrated to be extremely effective for Chern insulators \cite{Prodan2010ew} and the spin $\frac{1}{2}$ model \cite{ProdanJPhysA2010xk}. It involves an exact diagonalization of $H_\omega$ (and later of $P_\omega\hat{\sigma}P_\omega$) on a large lattice with periodic boundary conditions, and a large number ($10^3$) of disorder configurations. Energy levels are collected from a small window around a given energy $\epsilon$ and the level spacings between the collected levels are computed. By repeating this procedure for all disorder configurations, we generate an ensemble of level spacings which is statistically analyzed. We compute histograms showing the distribution of the level spacings and the variance $\langle s^2 \rangle$$-$$\langle s \rangle ^2$ for various energies $\epsilon$. The level spacing distributions are compared with the appropriate Wigner-Dyson surmise distributions, $P_{\mbox{\tiny{GUE}}}(s)$=$\frac{32 s^2}{\pi^2}e^{-\frac{4}{\pi}s^2}$ for the unitary case and $P_{\mbox{\tiny{GSE}}}$=$\frac{2^{18}}{3^6 \pi^3}s^4e^{-\frac{64}{9\pi}s^2}$ for the symplectic case, and to the Poisson distribution. If there is an agreement with the Wigner-Dyson surmise, which remains unchanged as the lattice size is increased, then one can safely conclude that the states at that energy are delocalized \cite{Evangelou1996yc,Cuevas1998xc}. In such cases, the variance of the level spacings converges to the theoretical variance of the Wigner-Dyson distributions, namely, 0.178 if the operator is in the unitary class and 0.104 if in the symplectic class. When the distribution of the level spacings agrees with the Poisson distribution and the variance is large, one can safely conclude that the states are localized. 

\begin{figure*}
\includegraphics[width=17.5cm]{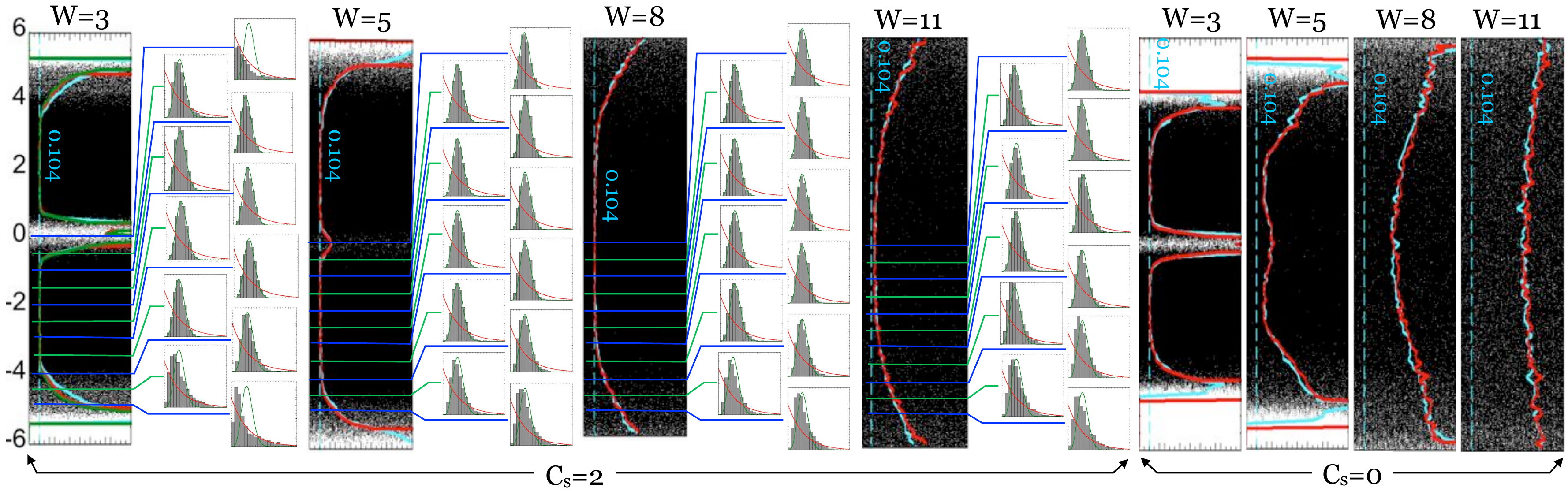}\\
\caption{The energy spectrum and level statistics at different disorder strengths, for the topological and trivial cases. The variance is plotted for increasing lattice sizes.}
\label{SpinChern2LevelStatistics}
\end{figure*}

Throughout this paper, we plot the spectrum of random operators as follows. For each disorder configuration, we place the set of eigenvalues on a vertical line and then we bring these lines near each other to generate plots like the ones seen in Fig.~\ref{SpinChern1LevelStatistics}. The representation is useful because one can see with naked eyes the regions where the eigenvalues display large fluctuations, such as the fuzzy regions near the edges of the spectra in Fig.~\ref{SpinChern1LevelStatistics}, which is an indicative of localization. Sometime one can also see the level repulsion (see Fig.~3 of Ref.~\cite{Prodan2010ew}), which is an indicative of delocalization. On top of the spectrum, we place the variance of the level spacings (always on a scale from 0 to 1)  so that we can corroborate its values with the fuzziness of  the spectrum. 

When decifering the data, it is instructive to start from the spin $\frac{1}{2}$ model, which was extensively analyzed in Ref.~\cite{ProdanJPhysA2010xk}. In the first panel of Fig.~\ref{SpinChern1LevelStatistics}, corresponding to a moderate disorder $W$=3, one can see energy regions where the variance is large but also two distinct regions where the variance becomes extremely close to 0.104. This feature was shown \cite{ProdanJPhysA2010xk} to remain unchanged when the lattice size is increased so one can safely conclude that these regions contain extended states. This finding is in line with previous studies based on transfer matrix analysis \cite{Onoda:2007xo}, Chern parity \cite{Essin:2007ij} or the non-commutative $Z_2$ invariant \cite{Loring2010xi},  and it is widely accepted nowadays. As the disorder is increased, the delocalized spectral regions in Fig.~\ref{SpinChern1LevelStatistics} don't suddenly disappear but instead they migrate towards each other until they meet and only then they disappear. This behavior is called levitation and pair annihilation and it is the hallmark of the extended states carrying a topological number. Such extended states cannot disappear unless they collide with other extended states carrying the opposite topological number, in which case the topological invariants cancel each other and the states become trivial and they immediately localize when the disorder is further increased. If we look instead at the trivial insulator in Fig.~\ref{SpinChern1LevelStatistics}, the behavior is completely different: the levitation and pair annihilation is absent and the spectrum becomes localized even at small disorder.   

The data for the spin $\frac{3}{2}$ model is presented in Fig.~\ref{SpinChern2LevelStatistics}. The calculations for this case are much more tedious because the dimension of the Hilbert space doubles when we go from spin $\frac{1}{2}$ to spin $\frac{3}{2}$. Consequently, we were able to complete the calculations only for 20$\times$20, 25$\times$25 and 30$\times$30 unit cells lattices. Even for these lattices, the computational effort was substantial, measuring in months of CPU time. Nevertheless, we saw a convergence with the lattice size so our conclusions are robust. The first panel of Fig.~\ref{SpinChern2LevelStatistics} corresponds to the topological case (whose clean limit was analyzed in Fig.~\ref{CleanSystem}(a-b)) with moderate disorder $W$=3. The histograms of the level spacing ensembles recorded at various energies are shown in the panels immediately to the right of the energy spectrum. The distributions are Poisson near the edges of the spectrum, while they become similar to $P_{\mbox{\tiny{GSE}}}(s)$ as we move towards the center of the bands. The variance, computed for the three lattice sizes mentioned above, has regions where it becomes practically equal to 0.104, the variance of $P_{\mbox{\tiny{GSE}}}(s)$. These features are stable as the size of the lattice is increased, hence we can safely infer the occurrence of delocalized bulk spectrum. As the disorder is increased, the delocalized spectral regions levitate until they touch and then disappear at around $W$=11, totally analogous to what we have seen in the spin $\frac{1}{2}$ calculation. There is only one possible conclusion: for the topological case, the spin $\frac{3}{2}$ model has protected extended states carrying a non-trivial topological invariant. If we examine the trivial case (whose clean limit was analyzed in Fig.~\ref{CleanSystem}(c-d)), we see that the levitation and pair annihilation is absent and instead the states localize even at moderate disorder, such as $W$=5.

We are going to show in the following that the bulk extended states seen for the spin $\frac{3}{2}$ model are protected by the spin-Chern number introduced in Ref.~\cite{Sheng:2006na}, using the new formulation given in Ref.~\cite{Prodan:2009oh}. Let $\hat{\sigma}$ be defined by $\hat{\sigma}|{\bm n},\sigma\rangle = \mbox{sgn}(\sigma) |{\bm n},\sigma\rangle $. In general, the spectrum of $P_\omega \hat{\sigma} P_\omega$ is symmetric relative to the origin with the positive and negative parts separated by a mobility gap \cite{ProdanJPhysA2010xk,Prodan:2010cz}. If $P^{\pm}_\omega$ define the projectors onto the positive/negative spectrum of $P_\omega \hat{\sigma} P_\omega$, then one can define corresponding Chern numbers:
\begin{equation}\label{Chern}
\begin{array}{c}
C_\pm=\frac{2\pi}{i} \left \langle \sum_\sigma \langle 0,\sigma|P_\omega^\pm \big{[} [\hat{x}_1,P_\omega^\pm],[\hat{x}_2,P_\omega^\pm] \big{]}|0,\sigma\rangle \right \rangle,
\end{array}
\end{equation}
via the non-commutative formula of Bellissard et al \cite{BELLISSARD:1994xj}. The outer angular parenthesis in Eq.~\ref{Chern} signify disorder average. Then the non-commutative spin-Chern number is defined as $C_s$=$\frac{1}{2}[C_+-C_-]$. 

\begin{figure*}
\includegraphics[width=15cm]{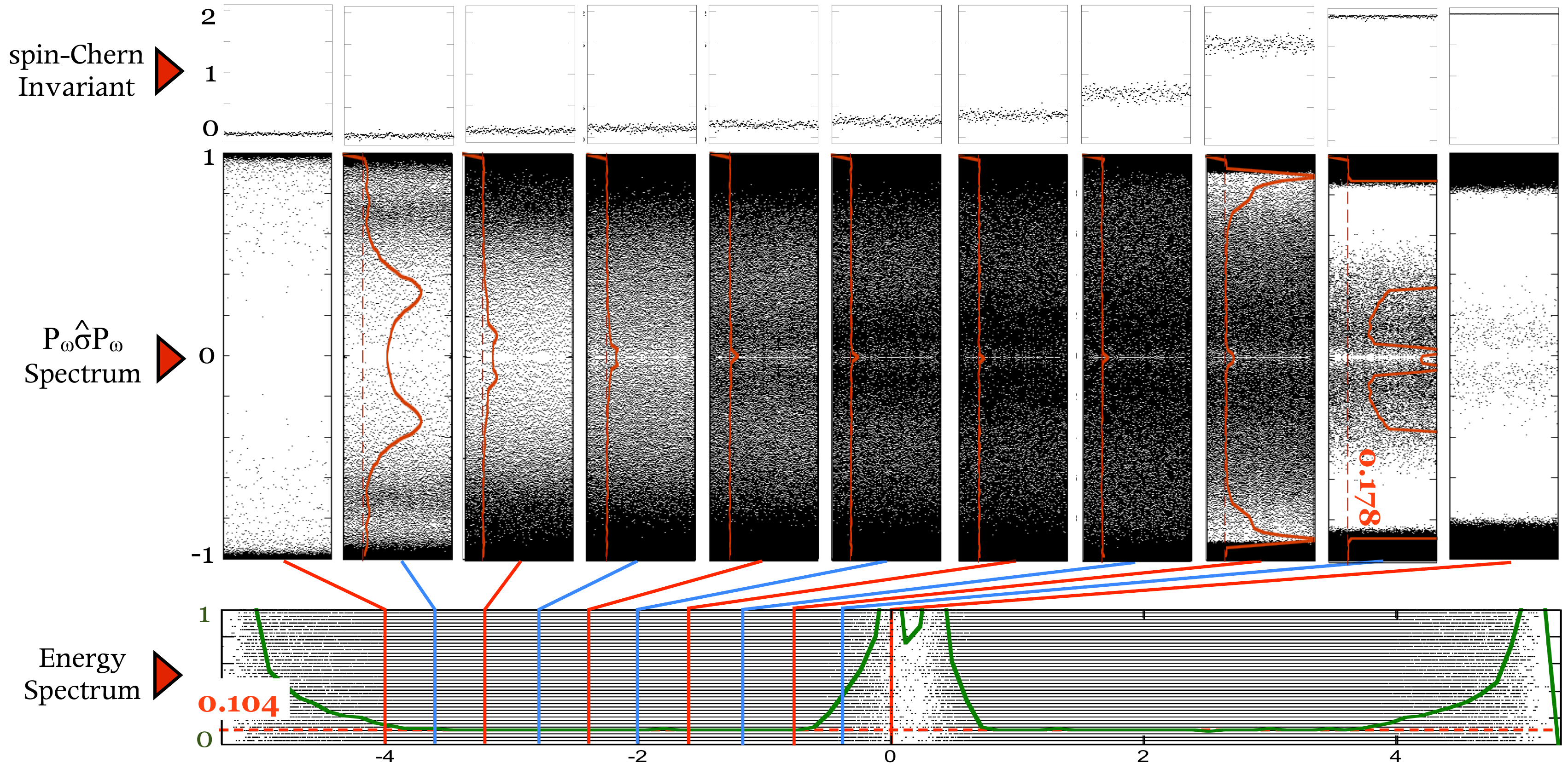}\\
\caption{Bottom: The spectrum of $H_\omega$ and the variance of the energy level spacings at $W$=3. Middle: The spectrum of $P_\omega \hat{\sigma} P_\omega$ and the variance of the level spacings for various Fermi levels. Upper: The numerical value of the spin-Chern number for the marked Fermi levels. The calculation was completed on a 30$\times$30 lattice.}
\label{SpinChern30x30W3}
\end{figure*}

In Ref.~\cite{Prodan:2009oh} it was realized that the projectors $P_\omega^\pm$ fit into the non-commutative theory of the Chern number \cite{BELLISSARD:1994xj} and consequently we can assert with absolute confidence that $C_\pm$, and therefore $C_s$, are quantized and invariant as long as $P_\omega^\pm$ remain localized. The delocalization of these projectors can happen via two mechanisms: 1) the mobility gap of $P_\omega \hat{\sigma} P_\omega$ closes, or 2) the projector $P_\omega$ itself becomes delocalized. For the spin $\frac{1}{2}$ model, the mobility gap of $P_\omega \hat{\sigma} P_\omega$ was found to be extremely robust and only the second mechanism was observed \cite{ProdanJPhysA2010xk}. We are going to examine the numerical data and use analytic arguments to show that same happens for the $\frac{3}{2}$ model. If that is the case, then a nonzero value of $C_s$ would necessarily imply the existence of extended bulk states. Indeed, imagine that we continuously lower the Fermi level $E_F$ until it reaches the bottom of the spectrum, where $P_\omega$=0 and consequently $C_s$=0. We can then see that $C_s$ changes its value during this process, so for some values of $E_F$ the  $P_\omega^\pm$ must became delocalized. Since the first mechanism is absent, this delocalization can come only from the delocalization of the whole projector $P_\omega$, which implies the existence of bulk extended states.

In Fig.~\ref{SpinChern30x30W3} we mapped $C_s$ and the spectrum of $P_\omega \hat{\sigma} P_\omega$ as the Fermi level was continuously lowered. We computed $C_s$ by evaluating Eq.~\ref{Chern} on a finite lattice with periodic boundary conditions by following the fast converging procedure introduced in Ref.~\cite{Prodan2010ew} and detailed in Ref.~\cite{ProdanJPhysA2010xk}.  We mention that none of the calculations presented here would have been possible without this procedure, which increases the accuracy and reduced the CPU time tremendously when compared with the traditional twisted boundary conditions calculations \cite{Essin:2007ij}. Let us concentrate on the spectrum of  $P_\omega \hat{\sigma} P_\omega$ first. This operator is in the unitary class, so if mechanism 1) would have taken place, one should see in the spectrum of $P_\omega \hat{\sigma} P_\omega$ a very narrow band of extended states moving towards the origin and closing the mobility gap. Such narrow bands of extended states would have had a distinct signature on the level statistics analysis \cite{Prodan2010ew} and the variance would have had sharp valleys where it abruptly converged to 0.178. These valleys should move towards the origin in order to close the mobility gap of  $P_\omega \hat{\sigma} P_\omega$. Clearly we don't see that in the data of Fig.~\ref{SpinChern30x30W3}. Instead, we see a sudden total collapse of the variance onto the value 0.178. This collapse happens immediately after the Fermi level enters the energy region where the variance of the energy level spacings becomes 0.104. Such collapsing behavior was previously seen in the entanglement spectrum of a Chern insulator \cite{Prodan2010ew} when the Fermi level crossed the Anderson transition point. The origin of that collapse is well understood now and comes from the delocalization of $P_\omega$. Similarly for the present data, the collapsing behavior can only come from the delocalization of $P_\omega$.

The data on the spin-Chern number strongly supports the above conclusion. As we lower the Fermi level, $C_s$ is seen to remain quantized at 2 until the Fermi level reaches the point where the variance of the energy level spacings becomes equal to 0.104, and where we infer that the protected extended spectrum starts. From there on, $C_s$ starts decreasing until $E_F$ exists the energy region of extended states, when $C_s$ sets to exactly the value 0. Putting all three data analysis together, the energy level statistics, the $P_\omega \hat{\sigma} P_\omega$ level statistics and the $C_s$ computation we have no alternative but to conclude that the spin $\frac{3}{2}$ model has protected extended bulk states and the protection is provided by the spin-Chern invariant. 

One question remains, why does the mobility gap of $P_\omega \hat{\sigma} P_\omega$ remain open? There are two reasons for this. First, the disorder is spin independent, so it has negligible direct effects on $P_\omega \hat{\sigma} P_\omega$. It is reasonably clear that the mobility gap of the Hamiltonian is more sensitive to disorder than the mobility gap of $P_\omega \hat{\sigma} P_\omega$ is. Second, the following identity:
\begin{equation}
2P_\omega^\pm=P_\omega \pm P_\omega \sigma P_\omega \big (1-(i[\sigma,P_\omega])^2\big )^{-\frac{1}{2}}
\end{equation}
shows that $P_\omega^\pm$ relate to the Green's function of the self-adjoint operator $i[\sigma,P_\omega]$, evaluated at $\pm 1$. The spectrum of $i[\sigma,P_\omega]$ is always inside the interval $[-1,1]$ so we are basically probing the edges of its spectrum. But as long as $P_\omega$ is localized,  $i[\sigma,P_\omega]$ is a random matrix with well behaved entries so the edges of its spectrum are expected to be localized. These two arguments show that mobility gap of $P_\omega \hat{\sigma} P_\omega$ is extremely robust. 

In conclusion, we have presented three independent numerical arguments all converging to one conclusion, that of existence of protected extended bulk states in a system with trivial $Z_2$ invariant. The protection mechanism steams from the insensitivity of the mobility gap of $P_\omega \hat{\sigma} P_\omega$ to the spin-independent disorder, fact that gives the spin-Chern invariant a status similar to that of the Chern invariant in Chern insulators. This protection is probably not universal, but it is clearly taking place in the particular model examined here and will very likely occur in many other models and real materials. In a good sense, our analysis shows how little we understand the disordered topological insulators and how much is there to explore.

\emph{Acknowledgements.} This research was supported by a Cottrell award from the Research Corporation for Science Advancement. H.S. is a Kressel fellow at Stern College for Women of Yeshiva University.


%

\end{document}